\documentclass[11pt]{article} % use larger type; default would be 10pt
\usepackage[round,authoryear]{natbib}
\usepackage[nottoc,notlof,notlot]{tocbibind} % Put the bibliography in the ToC

\usepackage{geometry} % to change the page dimensions
\usepackage{graphicx} % support the \includegraphics command and options

\usepackage{booktabs} % for much better looking tables
\usepackage{array} % for better arrays (eg matrices) in maths
\usepackage{paralist} % very flexible & customisable lists (eg. enumerate/itemize, etc.)
\usepackage{verbatim} % adds environment for commenting out blocks of text & for better verbatim
\usepackage{subfig} % make it possible to include more than one captioned figure/table in a single float
\usepackage{amssymb}
\usepackage{amsmath}
\usepackage{graphicx} % Allows including images
\usepackage{booktabs} %
\usepackage{fancyhdr} % This should be set AFTER setting up the page geometry
\usepackage{sectsty}
\allsectionsfont{\sffamily\mdseries\upshape} % (See the fntguide.pdf for font help)
\usepackage[nottoc,notlof,notlot]{tocbibind} % Put the bibliography in the ToC
\usepackage[titles,subfigure]{tocloft} % Alter the style of the Table of Contents
\newcommand{\Sb}{\mathbf{\Sigma}}
\newcommand{\Cb}{\mathbf{C}}
\newcommand{\Ab}{\mathbf{A}}
\newcommand{\yb}{\mathbf{y}}
\newcommand{\Qb}{\mathbf{Q}}
\newtheorem{lemma}{Lemma}

\newtheorem{proof}{Proof}

\title{Network estimation in State Space Model with L1-regularization constraint}
\author{Anani Lotsi, \qquad Ernst Wit}

\begin{document}
\maketitle
%\begin{frontmatter}

\begin{abstract}

Biological networks have arisen as an attractive paradigm of genomic science ever since the introduction of large scale genomic technologies which carried the promise of elucidating the relationship in functional genomics. Ever since, gene regulatory network or reverse engineer network has been the object for studies for many biological systems. Microarray technologies coupled with appropriate mathematical or statistical models have made it possible to identify dynamic regulatory networks or to measure time course of the expression level of many genes simultaneously. However one of the few limitations fall on the high-dimensional nature of such data coupled with the fact that these gene expression data are known to include some hidden process. In that regards, we are concerned with deriving
a method for inferring  a sparse dynamic network  in a high dimensional data setting. We assume that the observations are noisy measurements of gene expression in the form of mRNAs, whose dynamics can be described  by some unknown or hidden process. We build an input-dependent linear state space model from these hidden states and  demonstrate how an incorporated $L_{1}$ regularization constraint in an Expectation-Maximization (EM) algorithm can be used to reverse engineer transcriptional networks from gene expression profiling data. This corresponds to estimating the model interaction parameters. The proposed method is illustrated on  time-course microarray data obtained from a well  established T-cell data. At the optimum tuning parameters we found genes TRAF5, JUND, CDK4, CASP4, CD69, and  C3X1  to have higher number of inwards directed connections and FYB, CCNA2, AKT1 and CASP8 to be genes with higher number of outwards directed connections. We recommend these genes to be object for further investigation. Caspase 4 is also found to activate the expression of  JunD which in turn represses the cell cycle regulator CDC2.  R Computer source code is made available at our website at http://www.math.rug.nl/stat/Main/Software.

\end{abstract}

% \linenumbers

%% main text
\section{Introduction}
Reverse engineer transcriptional networks or modeling differential gene expression as a function of time has provided a new approach for biological research. Technology is now available to track the expression pattern of thousands of genes in a cell in a regulated fashion and to trace the interactions of many of the products of these genes \citep{Bower01}. However, the sheer dimensionality of all possible networks combined with the noisy nature of the observations and the complex structure of genomic regulation and signaling have meant that simply reading off a network from the data turned out somewhat optimistic. Instead, only statistical models of sufficient biological relevance are capable of discovering direct and indirect interactions between genes, proteins and metabolites. The last decade has seen an explosion of techniques to infer network structure from microarray data. Models have now been developed to capture how information is stored in DNA, transcribed to mRNA, translated to proteins and then from proteins structure to function. These models include Boolean networks based on Boolean logic \citep{Kauffman93,Dhaeseleer01082000} where each gene is assumed to be in one of two states``expressed'' or`` not expressed'', graphical Gaussian models \citep{Schfer005}, Dynamic Bayesian Networks (DBNs) \citep{Perrin03}, vector autoregressive models (VAR) \citep{Fujita07}, ordinary differential equation models (ODE) \citep{Quach01122007,Cao15072008} in which the state is a list of the concentrations of each chemical species and the concentrations are assumed to be continuous; they change over time according to differential equations, stochastic differential equations (SDE) \citep{Chen:2005:SDE:1181323.1181363}
and finally state space models \citep{ClaudiaRangel06122004,Beal05}.

Integrating these models in mainstream statistics is an exciting  challenge from a theoretical, computational, and applied perspective. Among the above mentioned network modeling techniques, ordinary differential equations (ODEs) have been established in recent years to model gene regulatory, or, more generally biochemical-networks, since they provide a detailed quantitative description of transcription regulatory network. On the downside, they are prone to a large number of model parameters and are not well suited to deal with noisy data. Current methods for estimating parameters in ODEs from noisy data are computationally intensive \citep{RSSB:RSSB610}.

In our work, we consider a  state space model (SSM) framework which consists of two different spaces, i.e protein space and mRNA space. A SSM is a special case of dynamic Bayesian networks (DBNs) and include hidden factors into the model, eg. genes of which the expression values are not measured. The SSM  or the standard or central linear Gaussian state space models  \citep{Fahrmeir99, Fahrmeir97}, in the context of time-series gene expression, assume that the observed time series data, $y_{t}$ represent a $p$-dimensional vector of gene expression observations of $p$ genes at time $t$.  $y_{t}$ themselves are generated from an underlying sequence of $k$ unobserved (hidden) state variables $\theta_{t}$ that evolve according to Markovian dynamics across successive time points.  In a essence the model consists of a linear observation equation in  states and is  supplemented by a linear transition equation.    A general linear SSM can be written as \citep{Koopman01},
\begin{equation}
\theta_{t}=F\theta_{t-1}+\eta_{t}.
\label{eq.2.1s}
\end{equation}

\begin{equation}
y_{t}=Z\theta_{t}+\xi_{t}.
\label{eq.2.2s}
\end{equation}

where $F$ and $Z$ represent the model coefficients of dimensions compatible with the matrix operations required in (\ref{eq.2.1s}) and  (\ref{eq.2.2s}). The two terms $\eta_{t}$ and $\xi_{t}$ are zero-mean independent system noise and measurement noise, respectively with

\begin{equation}
E(\eta_{t}\eta^{'}_{t})=Q, \qquad E(\xi_{t}\xi^{'}_{t})=R.
\label{eq.2.22s}
\end{equation}
Both $Q$ and $R $ are assumed to be diagonal in many practical applications.

Choosing SSM to model network kinetics has a number of advantages.  Most importantly, it allows the inclusion of hidden regulators which can either be unobserved gene expression values or transcription factors. The assumption of incompleteness of our data is quite realistic in the sense that in a microarray experiment, we usually do not observe protein concentrations together with mRNA concentrations due to the technical difficulty involved in performing such experiments. Thus we see the data as just noisy measurement of mRNA concentrations, whose dynamics can be described by some hidden process which involves protein transcription factors and mRNA concentrations.  
Another advantage of fitting SSM to the data stems from the fact that the variables of interest in the form of gene expression such as mRNA and protein transcription factors are seen as random variables, allowing the representation of some stochasticity, which could arise from either the measurement process  or  the nature of the biological process.

Several authors have exploited Kalman filtering \citep{Stoffer05, Meinhold83} and  SSM of gene expression and used them to reverse engineer transcriptional networks. To this effect, \citet{WU04}, in modeling gene regulatory networks, used a two step approach. In the first step, factor analysis is employed to estimate the state vector and the design matrix; the optimum dimension of the state vector $k$ was determined by minimum BIC. In the second step, the matrix representing protein-protein translation ($F$ in our equation \ref{eq.2.1s}) is estimated using least squares regression. \citet{ClaudiaRangel06122004} has applied SSM to T-cell activation data in which a bootstrap procedure was used to derive a classical confidence interval for parameters representing gene-gene interaction through a re-sampling technique. \citet{Beal05}  approached the problem of inferring the model structures of the SSM using variational approximations in the  Bayesian context through which a Variational Bayesian  treatment provides a novel way to learn model structure and to identify optimal dimensionality of the model. Recently, \citet{Bremer09} used SSM to rank observed genes in gene expression time series experiments according to their degree of regulation in a biological process. Their technique is based on Kalman smoothing and maximum likelihood estimation techniques to derive optimal estimates of the model parameters; however, little attention was paid to the dimension of the hidden state.

In microarray analysis, the  number of predictors (genes) to be analyzed far exceeds the number of observations ($p>>n$). Faced with such explosion of data, regularization has become an important ingredient and is fundamental to high-dimensional statistical modeling. The Lasso \citep{Tibshirani96} is one of the few methods for shrinkage and selection in regression analysis that incorporates an $l_{1}$ regularization constraint to yield a sparse solution.  A considerable amount of literature has been published on regularization methods in areas with large data sets such as genomics. These studies include the followings:
\begin{itemize}
	\item The regularization paths for the support-vector machine \citep{Hastie04}.
	\item  The elastic net \citep{zou05}; for applications with unknown groups of predictors and useful for situations where variables operate in correlated groups.
	\item $L_{1}$ regularization paths for generalized linear models \citep{park07} and
	\item   The graphical lasso  \citep{Fried08} for sparse covariance estimation and undirected graphs.
\end{itemize}

Our approach is based on penalized log-likelihood inference in the context of the Expectation-Maximization (EM) algorithm  \citep{Dempster97, Beal05, Ghahramani96} in a state space model. Stated differently, we fit a state space model (SSM) to the data and  demonstrate how an incorporated $L_{1}$ regularization constraint in EM algorithm can be used to reverse engineer transcriptional networks from gene expression profiling data.  State space models are good candidates to represent interactions between biological components in the form of mRNA concentrations and protein transcription factors. We present a statistical method that infers the complexity, the dependence structure of the network topology and the functional relationships between the genes, and deduce the kinetic structure of the network.  We estimate all model interaction parameters in order to clarify and describe the complex transcriptional response of a biological system and to clarify interactions between components.

Gene regulatory networks are usually sparse. Also, molecular ontologies suggest few connections among the many thousands of genes i.e, each gene may only be regulated by a few number of other genes or transcriptions factors . For that reasons, we will expect  many of the parameters to be zero leading to a sparse solution. It is in this context that we  employ a  regularization approach for the estimation of the parameters. This, form the basis for the $L_{1}$ penalization. The  proposed method in the maximization step of the EM-algorithm is the$L_{1}$ penalty through a simple modification of  the LARS algorithm by \citet{Efron04}, (Least Angle Regression). LARS is an efficient algorithm for computing the entire regularization path for the Lasso.

In this paper, we demonstrate how an incorporated $L_{1}$ regularization constraint in the EM algorithm  is  used in the  maximum likelihood set-up to reverse engineer transcriptional networks from gene expression profiling data. By so doing, we are able to add some useful interpretations to the model. We use the minimum AIC to determine the optimum level of sparsity.

The rest of this article is organized as follows. In section 2, we introduce the model, and give it a precise mathematical and biological interpretation. Section 3 describes the inference method and the model selection technique.  Section 4 is the application of our model to a real data (T-cell data) and summary of our results. We  conclude with a discussion of the method used, possible extension and a summary of related work in section 5.

\section{GENOMIC STATE SPACE MODEL}
We extend the model (Equations \ref{eq.2.1s} and \ref{eq.2.2s} ) by considering an input dependent SSM for gene expression times series data, where we allow inputs to both the state and observation equation as in  \citep{Beal05}.  The framework captures the stochastic nature of our biological process and their dynamics. To define the model, we start with the definition of the state variables $\theta_{t}$ represented as hidden process and the observation measurements are assumed to be produced by these hidden processes.  The model assumes that the evolution of the hidden variables $\theta_{t}$ is governed by the state dynamics which follows an input dependent first-order Markov process. The hidden variables $\theta_{t}$ are classically used to represent genes that have not been included in the microarray experiment, or unmeasured protein regulators, or transcription factors. The hidden variables are further corrupted by a Gaussian intrinsic biological noise $\eta_{tr}$. The hidden variables are not directly accessible but rather can be observed through  the observed data vector, $y_{t}$, namely the quantity of mRNA produced by the gene at time $t$. In essence we build a dynamic model that connects the observed variables $y_{t}$ (RNA transcripts) to the $k$-dimensional real valued unobserved quantities  $\theta_{t}$ such as unmeasured typically protein regulators.\par
Our model is defined through the following dynamics: 
\begin{itemize}
	\item First the state dynamics or the state of the network satisfies an input dependent first-order Markov process
 \begin{equation}
\theta_{tr}=F\theta_{t-1,r}+Ay_{t-1,r}+\eta_{tr}.
\label{eq.2.1}
\end{equation}
where $F$ is a regulatory matrix that quantifies the effect of  transcription factors at consecutive time points and is of dimension $k$ by $k$.  The quantity $A$ represents the translation matrix or input to the state matrix whose dimension is  $k$ by $p$,  $r=\left\{1,2,...,n_{R}\right\}$ denotes biological replicates and $\eta_{tr}$ is the Gaussian noise with mean $0$ and variance-covariance matrix $Q$. The initial state $\theta_{0}$ is  Gaussian distributed with mean $a_{0}=0$ and variance-covariance $Q_{0}$
 
\item Second the $p$ observation dynamics $y_{t}$ is a possibly time-dependent linear transformation of a $k$- dimensional real-valued $\theta_{t}$ with observational Gaussian noise $\xi_{t}$ and is given by 
 \begin{equation}
y_{tr}=Z\theta_{t,r}+By_{t-1,r}+\xi_{tr}.
\label{eq.2.2}
\end{equation}
where $Z$ describes how transcription factors regulate the transcription of genes with dimension  $k$ by $p$.  $B$ represents either  degradation or production matrix of mRNAs also known as input to observation matrix whose dimension is  $p$ by $p$ and $\xi_{tr}$ is the measurement Gaussian noise with mean $0$ and variance-covariance matrix $R$..
\end{itemize}

\begin{figure}[!ht]
	\centering
		\includegraphics[height = 5cm, width = 5cm]{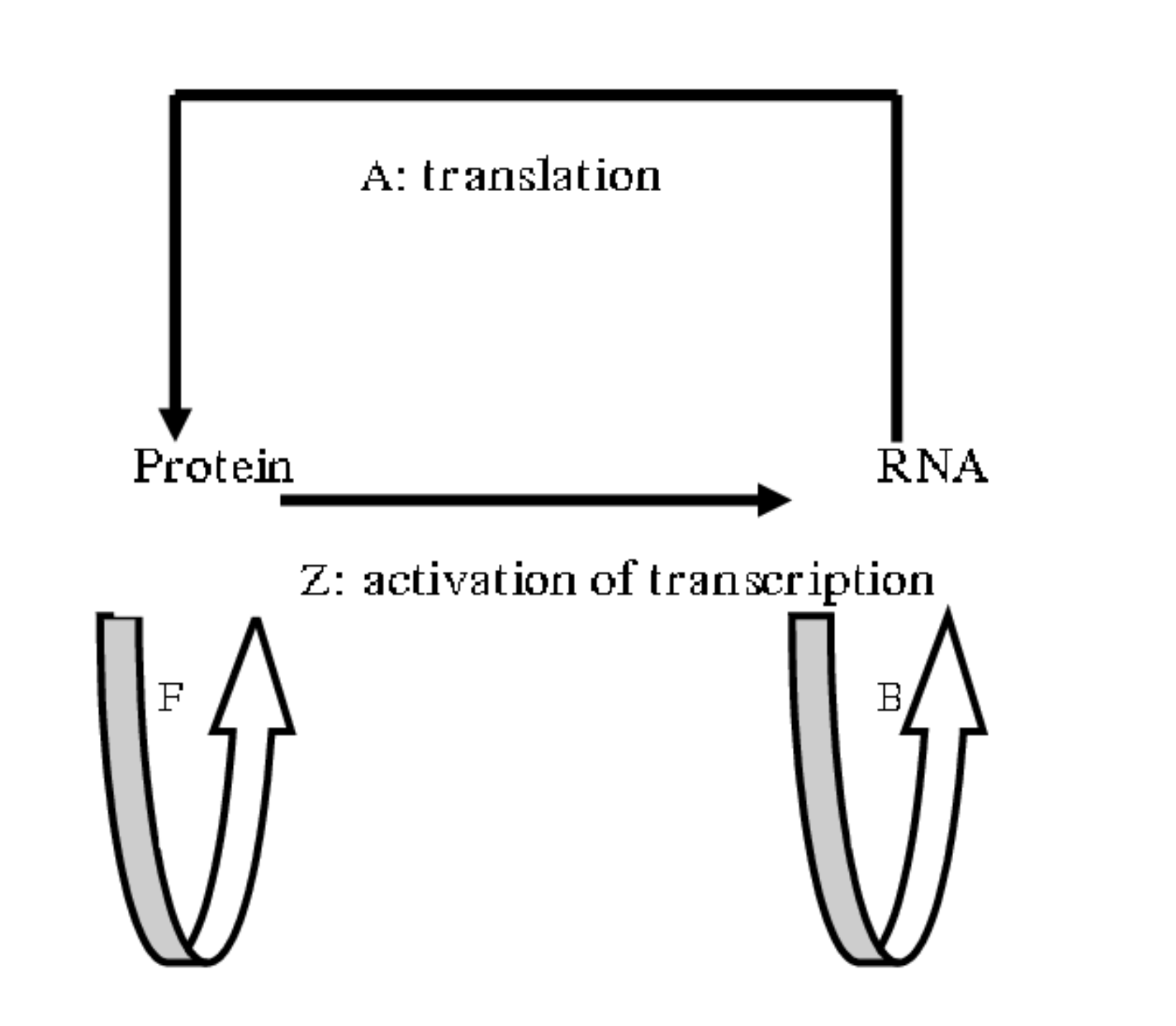}
	\caption{Biological interpretation of the input SSM.}
	\label{fig:dogma}
\end{figure}

The input dependent state space model defined by Equations (\ref{eq.2.1}) and (\ref{eq.2.2}) is an extension of the central SSM defined in Equations (\ref{eq.2.1s}) and (\ref{eq.2.2s}) and has been exploited in reverse-engineering transcriptional network \citep{Beal05, ClaudiaRangel06122004}.
The model indicates two scale networks, the protein space and the mRNAs space, across consecutive time points. It assumes  RNA-protein translation at two consecutive time points through the matrix $A$, and instantaneous protein-RNA transcription through $Z$. A biological  interpretation of the model network is also represented in Figure \ref{fig:dogma} which  describes two fundamental stages in gene regulation which are in conformity with the central dogma which states that DNA does not code for protein directly but rather acts through $2$ stages, namely, transcription and translation. The translation matrix $A$ also known as observation-to-state matrix  models the influence or the effects of the gene expression values from previous time steps on the hidden states. The matrix  $B$  indicates the direct gene-gene interactions. The state dynamic matrix $F$ describes the temporal development of the regulators or the evolution of the transcription factors from previous time step $t-1$ on the current time step $t$. It provides key information on the influences of the hidden regulators on each other. The observation dynamics matrix $Z$ relates the transcription factors to the RNAs at a given time point.
We now collect the model interaction parameters into a single vector $\varphi$ i.e $\varphi=\left\{G,Q,R,Q_{0}\right\}$ where $G=\left[\begin{array}{ccc}
B & Z&\\
A & F &\\
\end{array}\right]$ represents our graph of interactions. 
\section{LEARNING STATES AND PARAMETERS WITH RECURSIVE EM-ALGORITHM}

\subsection{Identifiability issues}

Briefly speaking, a parameter of a dynamic system is said to be identifiable given some data if only one value of this parameter can produce the observed likelihood. The identifiability property is  important because it guarantees that the model parameter can be determined uniquely from the available data. The poor identifiability of the SSM stems from the fact that  given the original model (Equations \ref{eq.2.1} and \ref{eq.2.2}), and with the linear transformation of the state vector $\theta^{*}_{t}=T\theta_{t}$, where $T$ is a non-singular matrix, we can find a different set of parameter vector say

 \begin{equation}
\mathbf{\hat{\varphi}}^{*}=\left\{\hat{G}^{*},\hat{Q}^{*},\hat{R}^{*}\right\}
\label{LN}
\end{equation} that give rise to the same observation sequence $\left\{y_{t},t=1,2,...,T\right\}$ having the same likelihood as the one generated by the parameter vector $\varphi$. Hence, if we place no constraints on $F$, $A$, $Z$, $B$ and possibly $Q$ and $R$, there exists an infinite space of equivalent solutions $\hat{\varphi}$ all with the same likelihood value.  To overcome such identifiability issues, further restrictions have to be imposed on the model.  In our work, we subject $Q$ and $R$ to be identity matrices with a careful choice of $Q_{0}$. Subjecting $Q$ to be identity only affects the scale of $\theta$  and matrices $A$ and $Z$.  The   matrices $A$ and $Z$ are then identifiable from the data, which can be seen from the marginal covariance matrix of $y$, $\Sb_{y}$. The latter, according to Schur complement \citep{Horn90} is given by  
\begin{equation}
\Sb_{y}=\left(K_{yy}-K_{y\theta}K^{-1}_{\theta\theta}K_{\theta y}\right)^{-1}
\label{eq.3.1}
\end{equation} 
 where $K_{yy}$ specifies the concentration matrix of the conditional statistics of the observed variables given the hidden variables and it is usually sparse and the quantity $K_{y\theta}K^{-1}_{\theta\theta}K_{\theta y}$ summarizes the effect of marginalization over the hidden variables and is of low rank; \citep{Pabl02010}. 

We further assume that the errors $\left\{\eta_{t}, t=1,...,T\right\}$ and $\left\{\xi_{t}, t=1,...,T\right\}$ are  uncorrelated. \par

\subsection{The likelihood function}
The  model interaction parameters are now restricted to  $\varphi=\left\{Z,B,F,A\right\}$ which is equivalent to our graph of interactions $G$. As can be seen from the model, the observations at time $t$, $y_{tr}$ are conditioned on the past observations, $y_{(t-1)r}$ and on the regulators $\theta_{tr}$ and also to infer for instance $\theta_{tr}$, we need $\theta_{(t-1)r}$ and $y_{(t-1)r}$. To that effect, under the Gaussian assumption we have the following:

\begin{eqnarray*}
\theta_{0r} & \sim &  N_{k}(\theta_{0r}|0,Q_{0}) \nonumber\\
\theta_{tr}|\theta_{(t-1)r},y_{(t-1)r} & \sim &   N_{k}(\theta_{tr}|\tilde{\theta}_{tr},Q)\nonumber\\
y_{tr}|\theta_{t},y_{(t-1)r}& \sim &   N_{p}(y_{tr}|\tilde{y}_{tr},R).\nonumber\\
\end{eqnarray*} where $$\tilde{\theta}_{tr}=F \theta_{t-1}r+Ay_{t-1}r,$$  $$\tilde{y}_{tr}=Z\theta_{tr}+By_{t-1}r,$$ and $N(.|\mu, \Sigma)$ is the normal density with mean $\mu$ and variance covariance matrix $\Sigma$.\par

We now write the marginal likelihood function $l^{m}_{y}(\varphi)$ of the data $y$. This is given by

\begin{eqnarray}
l^{m}_{y}(\varphi)&= & \int\prod^{T}_{t=1}P(\theta_{t}|F,A,\theta_{t-1},y_{t-1})P(y_{t}|B,Z,\theta_{t},y_{t-1})d\theta \nonumber\\
&= &\int \prod^{T}_{t=1}\psi(\theta_{t}|\tilde{\theta}_{t},\sigma^{2}_{\eta}I)\psi(y_{t}|\tilde{y}_{t},\sigma^{2}_{\xi}I)d\theta.
\label{eq.3.1m}
\end{eqnarray}  where $\psi(.|\mu,\Sigma)$ is the pdf of $N(.|\mu, \Sigma)$. The full  log-likelihood function of the complete data $(y_{tr},\theta_{tr})$ denoted by $l_{y,\theta}(\varphi)$ is for simplicity given by

\begin{equation}l_{y,\theta}(F,A,Z,B)=\sum^{n_{R}}_{r=1}l^{r}_{y_{r}\theta_{r}}(F,A,Z,B)
\label{eq.3.4}
\end{equation} where $l^{r}_{y_{r}\theta_{r}}(F,A,Z,B)$ is the complete log-likelihood of the $r^{th}$ replicate and is given by 
\begin{eqnarray}
l^{r}_{y_{r}\theta_{r}}(F,A,Z,B)& =&\sum^{T}_{t=1}l_{y_{t}|\theta_{t},y_{(t-1)}}(Z,B)+\nonumber\\
&& \sum^{T}_{t=1}l_{\theta_{t}|\theta_{(t-1)},y_{(t-1)}}(F,A).\nonumber\\
\label{eq.3.5}
\end{eqnarray}

From now onwards for a given replicate and for simplicity, we will write the unpenalized log-likelihood as:

\begin{equation}l_{y,\theta}(F,A,Z,B)=\sum^{n_{R}}_{r=1}l^{r}_{y_{r}\theta_{r}}(F,A)+\sum^{n_{R}}_{r=1}l^{r}_{y_{r}\theta_{r}}(Z,B)
\label{eq.3.8}
\end{equation} 

Learning the parameters of a state space model including the hidden variable can be tackled from different approaches. \citet{Beal05} solved the same problem using a Bayesian approach through a Variational Bayes  Method (VBM) that approximate the posterior quantities required for Bayesian learning.  As a probabilistic model, \citet{CRangel04} estimated the parameters trough a frequentist approach using maximum likelihood inference in the context of  EM algorithm. 
In our context, the number of parameters to be estimated  $P=k^2+2kp+p^2$ far exceeds  the number of observations. Thus we want to shrink unnecessary  coefficients to zero. This will make interpretation of results easier and probably reflects the true underlying situation by introducing some level of sparsity. To find the solution to the above problem, many well developed procedures can be used. For example, quadratic programming \citep{Tibshirani96}, the shooting algorithm \citep{Fu1998}, local quadratic approximation \citep{Fan2001} and most recently, the LARS method by \citet{Efron04} can all be employed. Our proposed method adapt the later procedure, optimization under $l_{1}$ constraint , where a penalty term is added to the likelihood function giving rise to a penalized likelihood criterion. LARS or optimization  with L1-regularization constraint turns out to be helpful and computationally feasible approach for finding sparse solutions in high dimension and by so rendering model interpretation easier. Therefore we now aim at maximizing the  marginal likelihood function $l^{m}_{y}(\varphi)$ of the data as given in \ref{eq.3.1m} subject to the constraints

\begin{equation}
||Z||_{1}\leq s_{1},\qquad ||B||_{1}\leq s_{2},\qquad ||A||_{1}\leq s_{3}, \qquad ||F||_{1}\leq s_{4}
\label{eq.3.1c}
\end{equation} where $s_{i}$ represents the regularization parameters or penalty parameters and we allow different  penalty parameters for different coefficients. Equations \ref{eq.3.1m} and \ref{eq.3.1c} are called  constrained regression problem. It is important to realize that the integration in \ref{eq.3.1m} involves the hidden component $\theta$, thus making the integration intractable.

\subsection{The EM algorithm}
As the true state variable is hidden, the integral in  \ref{eq.3.1m} is impossible. The  Expectation-Maximization (EM) algorithm stems from the fact that if we did have the complete data $(y_{t},\theta_{t})$ it will be straight forward to obtain maximum likelihood estimators (MLEs) of $\varphi$ using multivariate normal theory. In this case, we do not have the complete data. Therefore we use the EM algorithm. The latter is an  iterative method for finding the Maximum Likelihood Estimation (MLE) of   $\varphi$ using the observed data $y_{t}$, by successively maximizing  the conditional expectation of the complete data likelihood given the observed values.
The EM algorithm for SSM was formulated by \citet{Stoffer1982} and \citet{Shumway2000}. To this effect the algorithm requires the computation of the log-likelihood  of the complete data $l_{\theta,y}(\varphi)$  as given in Equation  \ref{eq.3.8}, and compute the conditional expectation of the log-likelihood given the  data. The algorithm is a two-stage procedure in which we begin with a set of trial initial values for the model parameter to calculate the Kalman smoothing $E(\theta_{r})$. The Kalman output are then input into the M-step to update parameter estimates subject to the constraints \ref{eq.3.1c}, giving rise to EM for penalized  likelihood estimation, \citep{Green90}.  The algorithm alternates recursively between an expectation step followed by a maximization step.

\subsubsection{The expected log-likelihood function: The E-step.}

 The E-step of the EM algorithm involves the calculation of the first two moments of $\theta$ of the hidden states i.e $E\theta$  and $E(\theta^{'}\theta)$.  Let $\Qb$ denote the expected log-likelihood. Then from Equation \ref{eq.3.8}, dropping the replicate index, $Q$  becomes
\begin{eqnarray}
\Qb(\varphi|\varphi^{*}) &= & E_{\theta,\varphi^{*}}\left[l_{y_{t},\theta_{t}}(\varphi)|\varphi^{*}y\right]\nonumber\\
&= &\sum^{T}_{t=1}E_{\theta,\varphi^{*}}\left[l_{y_{t},\theta_{t}}(\varphi)|\varphi^{*}y\right]\nonumber\\
&= &\sum^{T}_{t=1}E_{\theta,\varphi^{*}}\left[l_{y_{t}}(Z,B)|y\right]+\sum^{T}_{t=1}E_{\theta,\varphi^{*}}\left[l_{\theta_{t}}(F,A)|y\right]\nonumber\\
&= & Q_{1}(Z,B)+Q_{2}(F,A)
\label{eq.3.4*}
\end{eqnarray}
where 

	\begin{eqnarray}
\Qb_{1}(Z,B)&=&C_{1}+2\sum^{T}_{t=1}E(\theta^{'}_{t})Zy_{t}+2\sum^{T}_{t-1}y^{'}_{t-1}B^{'}y_{t}\nonumber\\
&-&\sum Z^{'}E(\theta^{'}_{t}\theta_{t})Z-2\sum^{T}_{t=1} E(\theta^{'}_{t})ZBy_{t-1}\nonumber\\
&-&\sum^{T}_{t=1} B^{'}y^{'}_{t-1}y_{t-1}B\nonumber\\
\label{eq.3.4h1}
\end{eqnarray}
and
	\begin{eqnarray}
\Qb_{2}(F,A)&=&C_{2}+2\sum^{T}_{t=1}y^{'}_{t-1}A^{'}E(\theta_{t})+2\sum^{T}_{t-1}E(\theta^{'}_{t})F E(\theta_{t-1})\nonumber\\
&-&\sum^{T}_{t=1} F^{'}E\theta^{'}_{t-1}\theta_{t-1}F-2\sum^{T}_{t=1}y^{'}_{t-1}A^{'}F E\theta_{t-1}\nonumber\\
&-&\sum^{T}_{t=1}A^{'}y^{'}_{t-1}y_{t-1}A\nonumber\\
\label{eq.3.4h2}
\end{eqnarray}
 $C_{1}$ and $C_{2}$ are known constants.
 
 The first two moments needed in the E-step are supplied by the Kalman smoothing algorithm  through a forward filtering pass and a backward smoothing pass, [see for instance \citep{Briers10}].
The above implies that for each replicate we run the Kalman smoothing algorithm to find the  expected hidden states and their variance-covariance components and these are joined together to get $\Qb(\varphi|\varphi^{*})$. 
Now Equation \ref{eq.3.4*} is the sum of two quadratic functions $\Qb_{1}$ and $\Qb_{2}$ that does not depend on $\theta$ but rather depend on the parameters and the data $y$ in a quadratic way. We maximize these two functions  during the maximization step.

\subsubsection{The update equations: The M-step.}
At this stage we solve for 

\begin{equation}
\varphi_{next}=argmax_{\varphi}\Qb(\varphi|\varphi^{*}) 
\end{equation} 
subject to the constraints defined in Equation \ref{eq.3.1c}. 

In essence we maximize iteratively the quadratic function $\Qb$ given in Equation \ref{eq.3.4*} across $\varphi$ using LARS algorithm, where each coefficient is assigned a tuning parameter $s$. This breaks down to two maximization problems, one for $\Qb_{1}$ across $(Z,B)$ and the other for $\Qb_{2}$ across $(F,A)$. The iterative maximization process is similar in both cases. 

We now show the maximization process for  $Q_{1}$. To do that, the following  lemma is needed. The proof is found in the appendix in the supplementary material.

\begin{lemma}
The solution that maximizes the quadratic function 
\begin{equation}
\Qb(X)=2b^{'}X-X^{'}SX \qquad  subject \quad to \quad ||X||_{1}\quad \leq s
\label{eq.3.4Q}
\end{equation}
is given by the lasso solution 

\begin{equation}
(\yb-\Cb \beta)^{'}(\yb-\Cb \beta) \qquad subject   \quad to \quad ||X||_{1}\quad \leq s 
\label{eq.3.4L}
\end{equation} where

\begin{equation}
\Cb=chol(S), \qquad \beta=Vec(X), \qquad \yb=\Cb S^{-1}b.
\end{equation} and the LARS solution of \quad(\ref{eq.3.4L})\quad is a function of

\quad  $S=\Cb^{'}\Cb$ \quad  and \quad $b=\Cb{'}\yb$; i.e,  \quad $f(\Cb^{'}\Cb,\Cb{'}\yb) $

\label{lem1}
\end{lemma}

Now from Equation \ref{eq.3.4h1}, given $B$ we now carry the maximization process of $\Qb_{1}$ across $Z$. Therefore, we can write $\Qb_{1}(Z,B)$ as  

\begin{equation}
\Qb_{1}(Z)=c_{1}+2b^{'}_{1}Z-Z^{'}S_{1}Z
\label{eq.3.4z}
\end{equation}
 subject to 
 \begin{equation}
 \sum^{kp}_{j=1} |z_{j}|\leq s_{2} 
\end{equation}
where $S_{1}=E\theta^{'}_{t}\theta_{t}$, $b_{1}=f(y,B,\theta)$, and $c_{1}$ is a constant.

Applying lemma \ref{lem1},  the update maximum likelihood estimators $\hat{Z}$  from Equation \ref{eq.3.4z} are just a function of $S_{1}$ and $b_{1}$. Therefore, we obtained the updates estimates $\hat{Z}$ by supplying the LARS function, the quantities $S_{1}$ and $b_{1}$ with a given tuning parameter where $b_{1}$  becomes the new data and  $S_{1}$ the data matrix. \par
Next, given $Z$, we maximize the quadratic function 

\begin{equation}
\Qb_{1}(B)=c_{2}+2b^{'}_{2}B-B^{'}S_{2}B
\label{eq.3.4b}
\end{equation}
 subject to 
 \begin{equation}
\sum^{p^2}_{j=1} |b_{j}|\leq s_{1}
\end{equation}
where $S_{2}=\sum^{T}_{t=1} y^{'}_{t-1}y_{t-1}$, $b_{2}=f(y,Z,\theta)$, $c_{2}$ is a constant. With the same analysis, the updates estimates $\hat{B}$ are obtained by supplying the LARS function, the quantities $S_{2}$ and $b_{2}$ with a given tuning parameter. Similar analysis is conducted for the estimation of $(F,A)$.

The advantage of this approach is that we see the LARS updates as functions not of the raw data, but instead as functions of  $S$ and $b$. This enables us to avoid first, the Cholesky decomposition of $S$ and second, computing $S^{-1}$ which are both time consuming and computationally inefficient.

\subsection{Model selection: Choice of regularization parameter $s$} \label{sel}
Determining the optimal SSM  tuning parameter $s$ is an important issue that deserves a thorough investigation. Popular  model selection criteria include the Mallow's $C_{p}$ \citep{Mallows73}, the  Akaike's Information Criterion (AIC) \citep{Akaike74} and the Bayesian Information Criterion (BIC) \citep{Schwarz78}. We apply Akaike's Information Criterion (AIC) method for our model selection. We generate a vector of values for the tuning parameters $s_{i}; i=1,...,4$. For each combination of the values of the tuning parameters we run the EM algorithm and obtain
$$   \hat{\varphi}(s)=argmax_{\varphi}\left[Q(\varphi)\right]$$  subject to constraints \ref{eq.3.1c}. \par

As recommended by \citet{Burnham02}, we have applied $AIC_{c}$ for our model selection procedure. The $AIC_{c}$ is given by

  \begin{equation} 
  AIC_{c}(\hat{\varphi}(s))=-2l(y_{t})+2P\left[\frac{N}{N-P-1} \right]
\end{equation} where $N=pTn_{R}$ represents total number of observations and $P=p^2+2kp+k^2$ is the total number of estimated parameters. Then for each model, the AICc is computed  and the model with the minimum AICc is selected. In essence,  minimizing AICc,  we obtained the optimal tuning parameters which is given by 
$$ \hat{s}=argmin_{s}AICc( \hat{\varphi}(s))$$
and the selected model parameters are given by $\hat{\varphi}(\hat{s}).$
Table \ref{Table II} summarizes the general formulation of the EM-$L_{1}$ penalized inference method:

\begin{table}[ht]
\begin{center}
\begin{tabular}{l}

  \hline \\
  
 1.  Iterate across penalty parameters $s \in S$ \\
	\quad	a. Start with initial values of  $\varphi$ \\

			\qquad i.  Do the E-step by calculating the Kalman smoother \\
		\qquad	ii. Perform the M-step via LARS algorithm \\

	\quad b. Repeat (i) and (ii) until convergence \\
			
 2.	Across $S$ select model with minimum AICc \\

   \hline\\
   
\end{tabular}
\caption{Summary of the EM for Penalized Likelihood inference method }
\label{Table II}
\end{center}
\end{table}

\section{Application}

For this study, to demonstrate the application of our reverse engineering  method, we used publicly available data, the results of two experiments used to investigate the expression response of human T-cells to PMA and ionomicin treatment. The data  is a combination of two data set namely tcell.34 and tcell.10. The first data set (tcell.34) contains the temporal expression levels of 58 genes for  $10$ unequally spaced time points.  At each time point there are 34 separate measurements. The second data set (tcell.10) come from a related experiment considering the same genes and identical time points, and contains 10 further measurements per time point. 

After pre-processing  the data, genes found to have few interactions were eliminated leaving us a total of 45 genes. At each time point there are 44 separate measurements or replicates. It was assumed that the 44 replicates have a similar underlying distribution. See \cite{ClaudiaRangel06122004} for more details. Given that the T-cell  is a time course gene expression data with technical replicates we expect more reliable estimation and inference results by applying our method.  Corresponding to each gene expression $y_{tr}$, we   also generated technical replicates  for the hidden variables $\theta_{tr}$.   In essence, we treated the data as a time series measurement data $y_{t_{r}}$, $t=1,2,...,10$ and $r=1,2,...,44$. Based on our previous work \citep{lotsi12a} and that of  \cite{Rau10}, we assumed the dimension of the hidden state $k$ to be $4$.  For each replicate, $y_{t}$ and  $\theta_{t}$ consist of $45$ genes and $4$ transcriptions factors respectively, each, measured at $10$ different time points, i.e for each replicate $r$, $y_{t}$ and $\theta_{t}$ are of dimension $(45\quad by\quad 10)$, $(4\quad by \quad10)$ respectively. Some of these genes include RB1, CCNG1, TRAF5, CLU.... The parameters $Q$ and $R$ were fixed and constrained to be identity.   We applied our $L_{1}$ penalized inference method to the time course gene expression data and estimated a total number of $2401$ parameters consisting of $B$, $A$, $Z$ and $F$. To do that, we iterate across the penalty parameters namely $4$ different tuning parameters $s_{B}$, $s_{A}$, $s_{Z}$ and $s_{F}$.  We started the EM- algorithm by taking as initial values of $\varphi$ say $\varphi_{0}$ the true parameters, and performed, the E-step to calculate the Kalman output. The latter became input for the M-step via LARS algorithm. While LARS produces the entire path of solutions, we make prediction or extract coefficients from the fitted LARS model using the ``predict'' function  in LARS. The predict function allows one to extract a prediction at a particular point along the path. This procedure is repeated until convergence. We then have different set of estimated model parameters corresponding to each set of tuning parameters. At this stage, we applied model selection technique via minimum AICc described in section \ref{sel} to select the optimum parameters. At the end, we  obtained the connectivity matrix of the directed genomic graph. The optimum estimated tuning parameters  has given rise to  fairly sparse networks.
\begin{figure}[!ht]
	\centering
		\includegraphics[height = 7cm, width = 8cm]{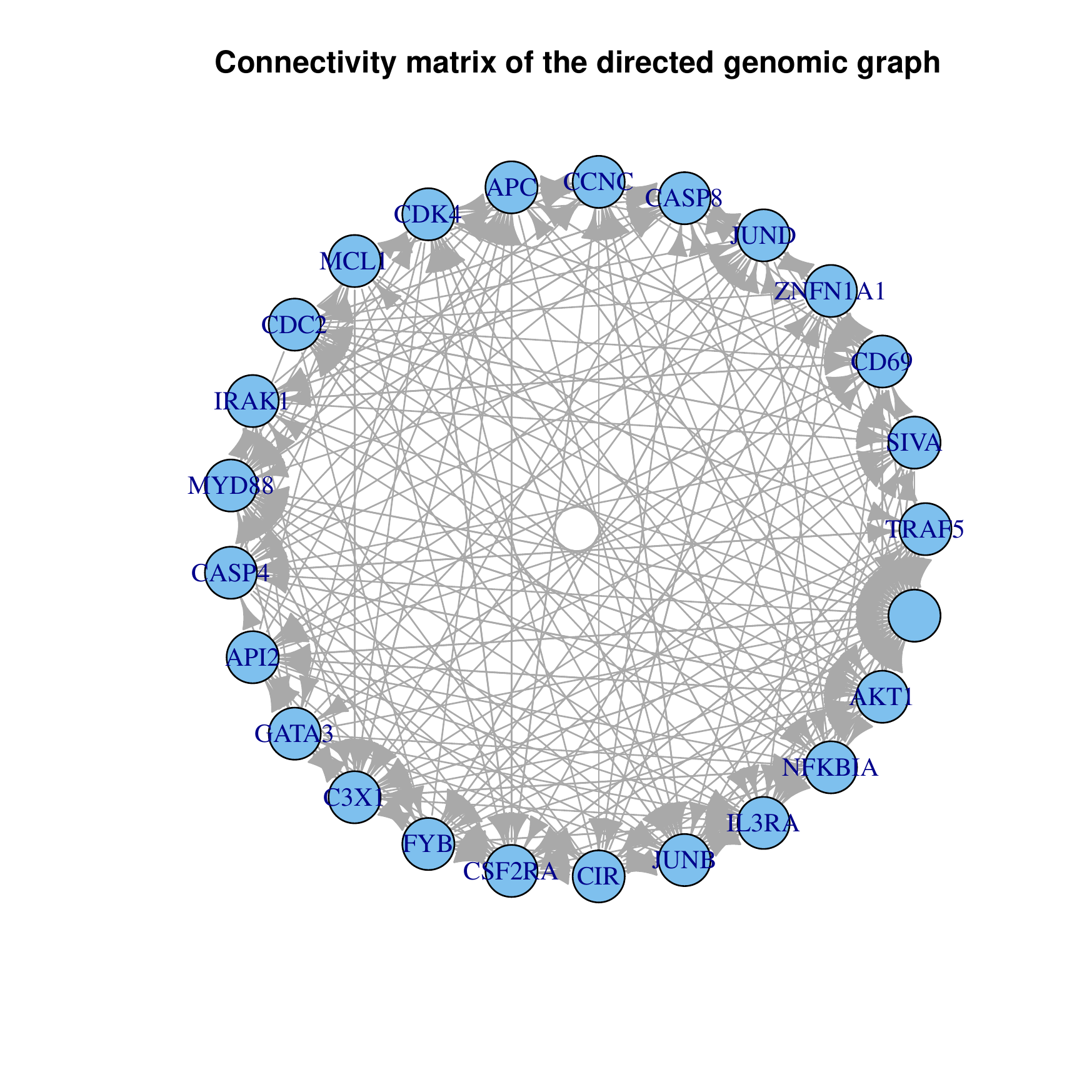}
	\caption{ Network found representing  the entire genomic interactions, $\mathbf{\hat{G}}$,  node refers to gene expression in the form of proteins or RNAs red solid lines and blue solid lines refer to inhibitory regulations and  activatory regulations respectively}
	\label{fig:fullnet}
\end{figure}

\begin{figure}[!ht]
	\centering
		\includegraphics[height = 7cm, width = 8cm]{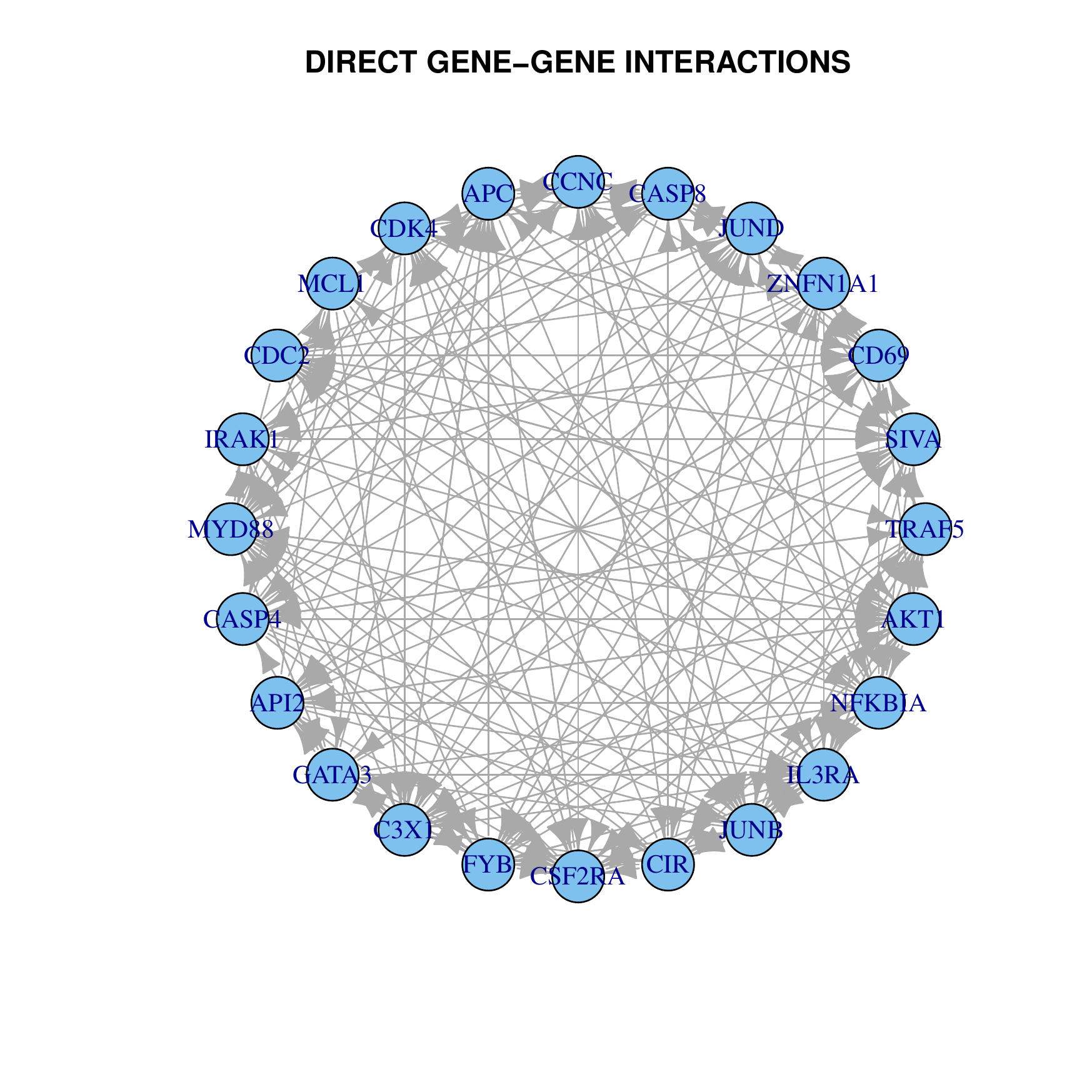}
		\caption{Network found representing  the gene-gene network interaction matrix, $B$, node refers to gene expression in the form  RNAs red solid lines and blue solid lines refer to inhibitory regulations and  activatory regulations respectively}
	\label{fig:RNAB}
\end{figure}

\begin{figure}[!ht]
\centering
		\includegraphics[height = 7cm, width = 8cm]{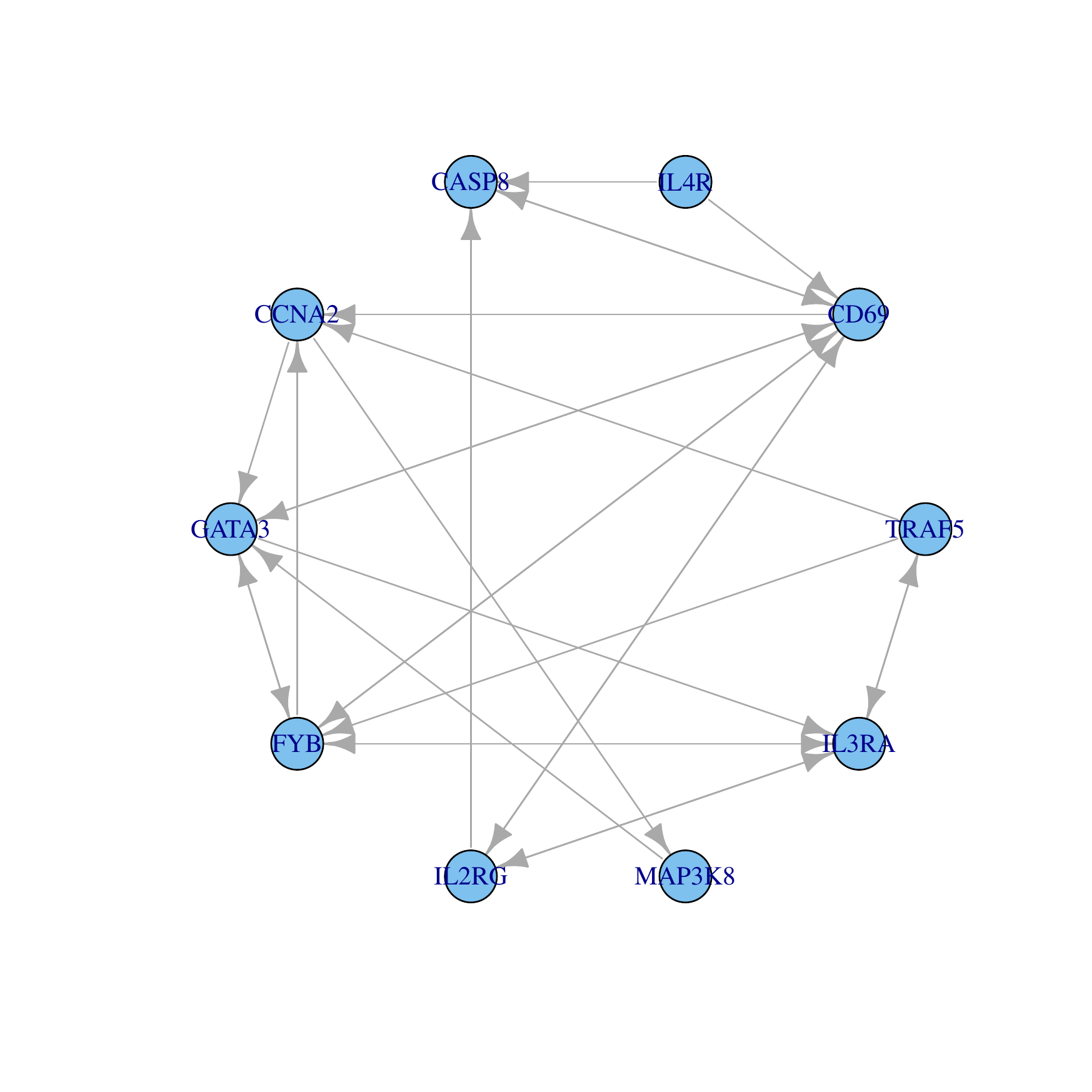}
		\caption{Subnetwork found representing the topology of gene FYB in connection with some selected genes}
\label{fig:subfyb2}
\end{figure}

The outputs are  graph showing connections from one gene expression variable at a given time point $t$ to another gene expression variable whose expression it influences at the next time point, $t+1$.
Figure \ref{fig:fullnet} depicts the graph of interaction $\mathbf{\hat{G}}$ i.e the entire genomic interaction representing mRNAs-mRNAs interactions $(B)$, proteins-proteins interaction $(F)$, proteins-mRNAs interaction $(Z)$ and finally mRNAs-proteins interactions $(A)$. Figure \ref{fig:RNAB} shows only the direct gene-gene interactions matrix $B$. The output depicted in Figure \ref{fig:subfyb2} is a subnetwork that  shows the topology of gene FYB. These $3$ figures shows that gene FYB is involved in the highest number of  outwards connections. The direction of the arrow indicates that  CCNA2, FYB, and CASP8 are mostly activation genes. Specifically,  FYB  activates  the expression of genes such as GATA3, CCNA2, CD69, IL3RA while CASP8 activates genes such as: JunD, CDC2, CD69. Figure \ref{fig:subnet} recovers the interactions between the Jun proteins family and other genes. It identifies Jun-D to have significant number of  connections in the form of activation and inhibitions. The structure of the network is visualized using the R package for Network analysis and visualization ``igraph".

\section{Results}
\begin{figure}[!ht]
	\centering
		\includegraphics[height = 7cm, width = 8cm]{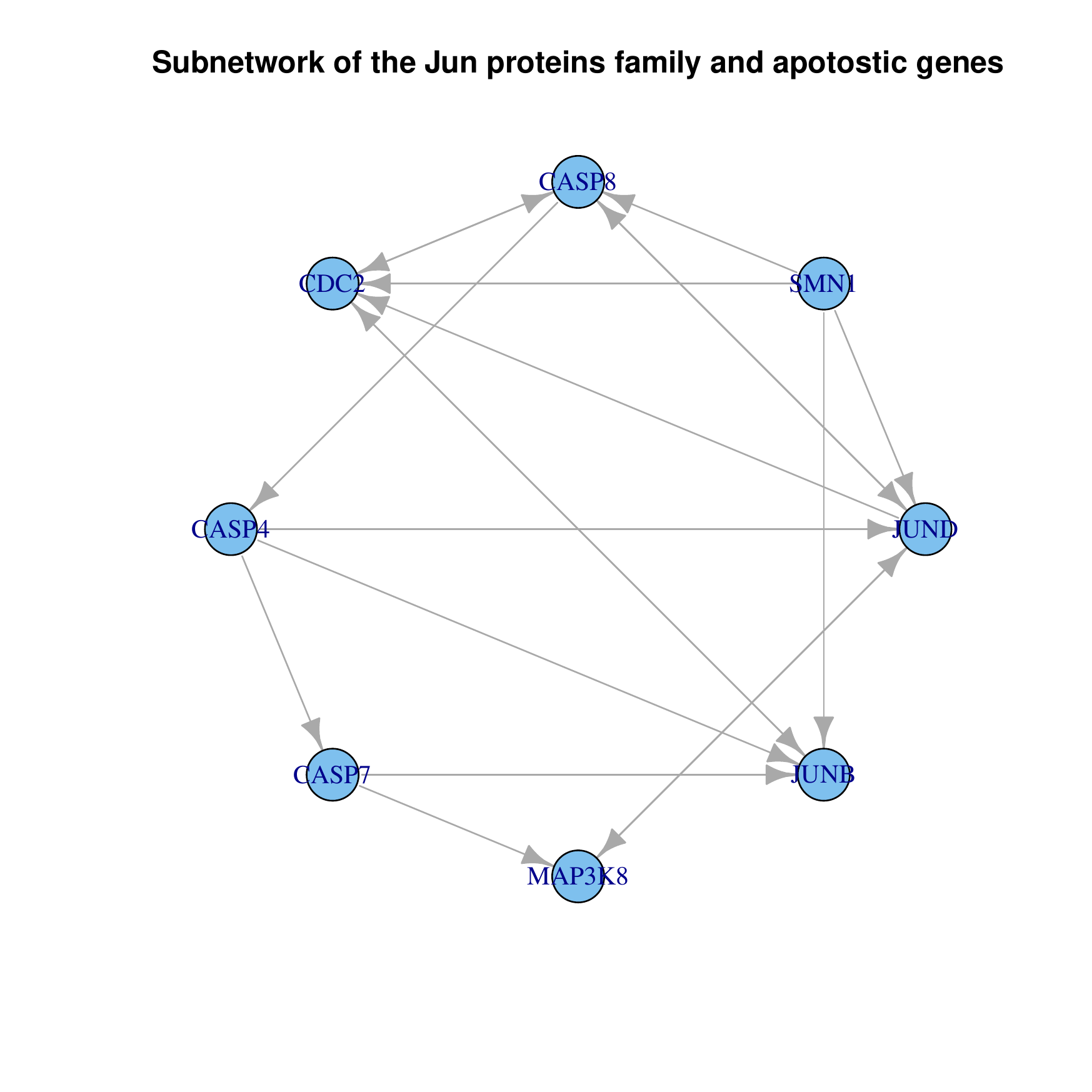}
		\caption{Subnetwork found representing the interactions between Jun proteins family and other genes}
	\label{fig:subnet}
\end{figure}

Our method has  resulted in  relatively  sparse networks as compared to  \citep{lotsi12a}. In all, the following genes were found to have the highest number of interactions in terms of inwards directed connections: TRAF5, C3X1, CASP4, CDK4 and  IL3RA. In addition, from a topological point of view genes such as JUND, AKT1, FYB, CCNA2 and  occupy a crucial position in the recovered networks. We recommend these genes to be object of further study by biologist. These results supports the works of \citet{ClaudiaRangel06122004, CRangel04}. Both  found gene FYB to occupy an important position in their respective graphs. At the optimum turning parameter, we found Jun-B interacting directly with  CASP4 through Jun-D; a result also supported by  \citet{Beal05} .  The unpenalized inference \citep{lotsi12a} approach has indicated that Jun B  activates  directly  CDC2. This work also supports the same interaction. A portion of the sub-networks found by  \citet{Beal05} and \citet{Rau10} representing the interactions between CASP4 and Jun-D is also found in our network through figure \ref{fig:subnet}. Another interesting interactions that was supported by previous literature were interactions between JUN-D and CASP7 on one hand and interactions between JUN-D and CDC2 both in the form of inhibitions.  JUN-D is predicted to repress the expression level of the cell cycle regulator CDC2.  This clearly supports the hypothesis that JUN-D negatively regulate cell growth and acts as anti-proliferative and anti-apoptotic signal gene. 

A critical look at figure \ref{fig:RNAB} reveals that AKT1 and MLC1 also occupy a crucial position.  AKT1 is  found to influence the expression level of many genes. Of these,  include the JUN proteins JUN-B , one interleukin  receptor gene (IL3RA), one apoptosis-related cysteine protease ( CASP4), the cell division cycle ( CDC2).  AKT1 is also seen to activate  the expression level  of a transcription factor.  In our model, MLC1 is found to regulate positively the expression level of one of the transcription factors. Also  MLC1 represses  several genes including  CASP7, CDK4, C3X1 to mention but few.  \citet{Rau10} found that CAPS4 inhibits the expression level of CAPS8. This interaction was  supported by our work.  \citet{Beal05} did not recover such interaction. We are constrained by the space limitation to provide details for all the biological findings in this methodological paper. Also some findings (Figure \ref{fig:subnet}) of our current study do not support the work of \citet{Beal05} in the sense that we found no interactions between Jun-D and Caspase-7 and also between JUN-B and MAPK8. Thus, the results based on our methodology suggest some findings that are supported by the current literature and are biologically interpretable, while some other findings have not been documented yet in biological literature and we hope these new findings will be confirmed in the near future. A comparison of our proposed model and method to alternative models and methods for dynamic network construction would be desirable, but is beyond the scope of this article.

\section{Conclusion}
In this paper, we have inferred a sparse dynamic network by fitting an input dependent linear state space model to T-cell data. We have assumed that the true biological process  is not fully observed and the hidden variables were first calculated using a Kalman filtering and smoothing algorithm via an E-step. We then proceed to update the model parameters through an $L_{1}$ regularization constraint via LARS algorithm in the maximization step. We used AICc to determine the optimum combination of tuning parameters and hence the model parameters. The method we presented in our paper can be viewed as an Expectation-Regularization-Maximization  approach which  produces a sparse high-dimensional gene dynamic regulatory network.  The proposed method offers significant advantages over other methods  that have recently appeared in the literature. For example,  \citet{Beal05} inferred  regulatory interactions from expression data by maximizing the marginal likelihood with a modification of the EM algorithm. His approach was based on variational Bayesian methodology which is an approximation of the posterior distribution of the parameters while we did exact inference of the parameters.  \citet{ClaudiaRangel06122004} used cross validation as model selection technique which is quite slow as compared to AIC.  Also our model allows for dynamic correlation over time, as each observation and hidden state depend explicitly on some function of previous observations  as opposed to the models described by  \citet{yamaguchi06, Perrin03, WU04}. Their models do not allow for RNA-protein translation (the matrix $A$ ) and RNA-RNA  interactions (the matrix $B$) in our model. Most importantly, the LARS algorithm adopted guarantees us interpretable model, and accurate predictors.

One fundamental assumption in our proposed model is the first-order linear dynamics in the state and observation equations of the SSM. The advantages of using linear SSM stems from the fact that the linearity assumption has resulted in  a more stable network and has enabled us to  recover the  dynamics of the network easily as compared to the nonlinear relationships. The inference method via LARS  is potentially revolutionary, offering interpretable models, relative stability, accurate predictions, unbiased inferences, and a nice graphical display of coefficient paths that indicates the key tradeoff in model complexity.  We used the R package for Network analysis and visualization ``igraph"to display  simple, and easy to understand graph through which the whole system under study can be ascertained quite easily. Gene regulatory interactions surely include complex interactions which nonlinear SSM may capture well. To recover the network from the nonlinear model  is however complicated in both theory and  computationally, especially in high dimensional setting.  Future works will encompass  extending the linear SSM into a  Non Linear State Space Model (NLSSM) \citep{Quach01122007} whose hidden process will be defined through an integration of an Ordinary Differential Equation (ODE) and estimate both parameters and hidden variables through the same inference technique. We also plan to overcome the ODE limitations namely ability to handle noisy data and the high number of model parameters by integrating a sparse ODE model into a graphical model framework, thus taking noisy measurement into account, and the resulting model will then be embedded into a penalized maximum likelihood learning set-up.

\section{Appendix}
We outline here the proof of  lemma 1 

\begin{proof}

Properties of Gaussian distribution and Gaussian processes suggest that the quadratic $Q(X)$
corresponds to a Gaussian $N(S^{-1}b, S^{-1})$. Therefore

\begin{eqnarray}
Q(\beta)& = &(\beta-S^{-1}b)^{'}\Sigma^{-1}(\beta-S^{-1}b)\nonumber\\
& = & (\beta-S^{-1}b)^{'}S(\beta-S^{-1}b)\nonumber\\
& = & (\beta-S^{-1}b)^{'}C^{'}C(\beta-S^{-1}b)\nonumber\\
& = & (\Cb S^{-1}b-\Cb \beta)^{'}(\Cb S^{-1}b-\Cb \beta)\nonumber\\
& = &(\yb-\Cb \beta)^{'}(\yb-\Cb \beta).\nonumber\\
\label{eq.3.4qx}
\end{eqnarray}
The second part of the proof follows  from Efron et al., page 7, Formula (2.4)-(2.13) which  suggest that the next step of LARS algorithm updates the coefficient $\hat{\beta}_{k-1}$, say to

\begin{equation}
\hat{\beta}_{k}=\hat{\beta}_{k-1}+\hat{\gamma}w
\label{eq.3.5p}
\end{equation}
where $w$ is the unit vector making equal angles, less than $90^{0}$ with the active columns of $C$, and $\hat{\gamma}$ is the step size.
For $k=0$, we assume $\hat{\beta}_{k}=0$. Now suppose that  $\hat{\beta}_{k}$ as defined above is a function of $C^{'}C$ \quad and \quad $C^{'}\yb$; i.e f($C^{'}C$,$C^{'}\yb$). 
We will show that the next coefficient
\begin{equation}
\hat{\beta}_{k+1}=\hat{\beta}_{k}+\hat{\gamma}w.
\end{equation}

is also a function $C^{'}C$ \quad and \quad $C^{'}\yb$

Define the followings:
\begin{itemize}
	\item $w=A*G_{i1}$ where \newline
				$A=\left[1^{'}_{\Ab}(\Cb^{'}_{\Ab}\Cb_{\Ab})^{-1}1_{\Ab}\right]^{-\frac{1}{2}}$. \newline
				$G_{i1}= ((\Cb^{'}_{\Ab}\Cb_{\Ab})^{-1})$  and $\Ab$ is a the subset of indexes of the active set; clearly $w$ is a function of $(\Cb^{'}\Cb)$. i.e
				
				\begin{equation}
				w=f(\Cb^{'}\Cb).
				\label{eq.3.5*}
				\end{equation}

  \item $\hat{\gamma}=min^{+}\left\{\frac{(C_{max}-c)}{A-a},\frac{(C_{max}+c)}{A+a}\right\}$ where  \newline
  $c=\Cb^{'}\yb$, representing vector of current correlations, $C_{max}$ is the maximum absolute value from the set $c$ \newline
  $a=(\Cb_{A^{c}})^{'}\Cb_{A}w$  \newline
 The above definitions of $c$ and $a$  indicate  that 
 
 	\begin{equation}
 	\hat{\gamma} = f(\Cb^{'}\Cb,\Cb^{'}\yb). 
 		\label{eq.3.5**}
				\end{equation}
  
\end{itemize}

Combining Equations \ref{eq.3.5*} and 	\ref{eq.3.5**}, we have that $\hat{\beta}_{k+1}$ is a function of $\Cb^{'}\Cb$ \quad and \quad $\Cb^{'}\yb$; i.e

	\begin{equation}
				\hat{\beta}_{k+1}=f(C^{'}C,C^{'}\yb) 
				\label{eq.3.5l}
				\end{equation}
				
\end{proof}	

\section{R code files}

The R code files run on the file "main.r".  The file main contains the following four major functions (files)
\begin{itemize}
	\item "funclibrary.r": This function contains all the library functions.
	\item "funcall.r": This function contains functions needed for the implementation of the EM algorithm
	\item "EM.r":  This function performs the EM algorithm
	\item "plotgraph.r": This function displays the networks or graph of interactions. 
\end{itemize}
N.B: What to download? 
\begin{itemize}
	\item Download all the sub functions  
	\item Run the File "main" on R environment. 
\end{itemize}

\renewcommand{\bibname}{Bibliography}
\bibliographystyle{mykluwer}
\bibliography{reference}
\end{document}